\documentclass[showpacs,pre]{revtex4}
\usepackage{amsmath}
\usepackage{graphicx}
\begin{document}
\title{Reaction-diffusion-advection equation in binary tree networks and optimal size ratio}
\author{Hidetsugu Sakaguchi}
\affiliation{Department of Applied Science for Electronics and Materials,
Interdisciplinary Graduate School of Engineering Sciences, Kyushu
University, Kasuga, Fukuoka 816-8580, Japan}
\begin{abstract}
A simple reaction-diffusion-advection equation is proposed in a dichotomous tree network to discuss an optimal network.  An optimal size ratio $r$ is evaluated by the principle of maximization of total reaction rate. In the case of reaction-limited conditions, the optimal ratio can be larger than $(1/2)^{1/3}$ for a fixed value of branching number $n$, which is consistent with observations in mammalian lungs. We find furthermore that there is an optimal branching number $n_c$ when the Peclet number is large. Under the doubly optimal conditions with respect to the size ratio and branching number, the optimal value of $r$ is close to $(1/2)^{1/3}$.
\end{abstract}
\pacs{05.45.Df, 05.60.Cd, 87.19.ug}
\maketitle
The problem of branching networks appears in many physical and biological systems~\cite{rf:1}. We consider biological tree networks such as blood vessels, bronchial trees, and plant vascular systems. The purpose of these systems is to transport materials. For example, oxygen is transported from the nose to the lungs via bronchial trees, and is further transported from the lungs to the whole body by the vascular plexus. The vascular plexus and bronchial networks have been studied from a medical point of view, because many diseases are related to malfunctions in these networks. Owing to the biological importance of these tree networks, these systems have also been studied physically or mathematically by many authors. Several authors studied the geometry of these tree networks. Murray studied the branching system from a parent blood vessel to two daughter blood vessels  by the principle of minimization of work, assuming stationary Poiseuille flow~\cite{rf:2}. The size ratio of parent and daughter branches was estimated at $(1/2)^{1/3}\sim 0.794$~\cite{rf:3,rf:4}. 
If self-similar dichotomous branching is assumed, the fractal dimension is $\ln 2/\ln (1/2^{1/3})=3$, that is, the tree network is space filling. Direct observations of these biological networks have also been carried out. The morphometry of the human lung was summarized by Weibel~\cite{rf:5}. The bronchial trees of human lungs have a clear dichotomous branching structure of 23 generations, and the ratio of duct sizes in the bronchial trees is estimated to be 0.85. The reason why the ratio is slightly larger than the theoretical value of 0.794 proposed by Murray is not well understood.  Kitaoka et al. constructed a detailed three dimensional model of the human airway tree~[6].  Mauroy et al. numerically studied the three-dimensional flow distribution in a branching tree, not assuming the stationary Poiseuille flow~\cite{rf:7}.   Diffusion, rather than convection,  becomes the dominant mode of transport of oxygen when the branching generation is larger than around 17 in human lungs. This is because the flow velocity becomes increasingly smaller in deeper branches.  The diffusion process at the deeper branches was also studied using fractal geometries~\cite{rf:8,rf:9}.

Although Murray et al. evaluated the optimal size ratio $r=(1/2)^{1/3}$ for a fixed branching generation number $n$, the work or power loss increases with $n$. This is because the flow resistance increases with $n$. The complicatedly-branched network is not suitable to transport a large amount of material. Then, why are biological networks so complicatedly branched? In this paper, we consider a reaction-diffusion-advection model in a dichotomous tree and propose a new principle of maximization of total reaction rate at the deepest branch tips. The number of deepest branch tips increases with $n$ and the total reaction rate can be facilitated in the complicatedly-branched network.  We keep the alveoli in bronchial trees in mind as an example of such deepest branch tips, although our model is too simple to apply directly to the bronchial trees. 

We consider a simple system of dichotomous branching ducts as shown in Fig.~1.
The radius and length of the first generation duct are set equal to $a$ and $b$. For the sake of simplicity, the size ratio of the parent and daughter ducts is assumed to take the same value  $r$.
The radius $a_i$ and length $b_i$ of the $n$th generation duct are therefore $ar^{n-1}$
 and $br^{n-1}$. 
The total volume $V_n$ from the first to the $n$th generation ducts is 
\begin{equation}
V_n=\pi a^2b\sum_{i=1}^{n}2^{i-1}r^{3(i-1)}=\pi a^2b\frac{1-(2r^3)^{n}}{1-2r^3}.
\end{equation}
We assume  Poiseuille flow in the circular duct, in which the flow is axisymmetric and laminar. A certain pressure difference is applied between the inlet and outlet of the duct. Then, the velocity $v_i$ in the $i$th generation duct is expressed as 
\begin{equation}
v_i=\frac{a^2r^{2(i-1)}\Delta P_i}{4\eta br^{i-1}},
\end{equation}
where $\eta$ is the viscosity and $\Delta P_i$ is the pressure difference between the inlet and the outlet of the $i$th duct. Owing to law of conservation of flow, which states that $2^{i-1}\pi a^2r^{2(i-1)}v_i=$const., the velocity $v_i$ is expressed as $v_i=1/(2r^2)^{i-1}v_1$. The total pressure difference $P$ between the inlet of the first generation duct and the outlet of the $n$th generation duct is expressed as
\begin{equation}
P=\frac{4\eta bv_1}{a^2}\sum_{i=1}^{n}\left(\frac{1}{2r^3}\right)^{i-1}=\frac{4\eta bv_1}{a^2}\frac{1-1/(2r^3)^{n}}{1-1/(2r^3)}.
\end{equation}
The flow rate $Q$ is expressed as $(1/2)\pi a^2v_1$ for Poiseuille flow in the first duct. From Eq.~(3), $Q$ is rewritten as 
\begin{equation}
Q=\frac{\pi a^4}{8\eta b}\frac{1-1/(2r^3)}{1-1/(2r^3)^{n}}P. 
\end{equation}
It can be shown that the principle of minimization of total energy dissipation is equivalent to the maximization of the flow rate $Q$ under the condition of fixed $V_n$. 
Figure 1(b) shows the relation between $Q$ and $r$ for $n=23$. The values of the other parameters are set to  $a=b=\eta=1$ and $V_n=23$. The flow rate $Q$ reaches its maximum value at $r=(1/2)^{1/3}\sim 0.794$. This is Murray's law for a fixed value of $n$. We can change the value of $n$ as a control parameter.   
Figure 1(c) shows the flow rate $Q$ at $r=(1/2)^{1/3}$ as a function of generation number $n$ in a double-logarithmic plot. The dashed line denotes $Q\propto 1/n^2$. 
At $r=(1/2)^{1/3}$, $V_n=n\pi a^2b$ and the flow $Q=\pi a^4 P/(8n\eta b)$.
Under the condition that $V_n$ and the ratio $b/a$ are constant, $a^3\propto V_n/n$, that is,  $a$ and $b$ change with $n$ as $a\propto 1/n^{1/3}$ and $b\propto 1/n^{1/3}$. Therefore, $Q\propto 1/n^2$.  This result implies that the flow rate $Q$ is maximum at $n=1$ and the efficiency of convective transport decreases with $n$. 
\begin{figure}[t]
\begin{center}
\includegraphics[height=4.cm]{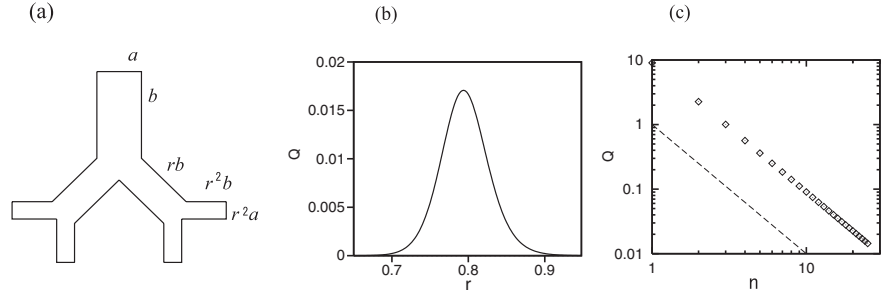}
\end{center}
\caption{(a) Dichotomous branching ducts. (b) Relation of the flow rate $Q$ and contraction rate $r$ for $n=23,a=b=\eta=1$ and $V_n=23$. (c) Relation of the flow rate $Q$ and total branching number $n$ for $a=b=\eta=1$ and $V_n=23$.}
\label{f1}
\end{figure}

The purpose of tree networks is not the transport of material but the reaction of material at branch tips.  For example, the purpose of bronchial tree networks is to exchange oxygen and carbon dioxide at the alveoli and capillary vessels.  The material is transported both by Poiseuille flow and by diffusion. For the transport of the material, we assume a one-dimensional diffusion-advection equation for the concentration $c$ in the $i$th generation duct:
\begin{equation}
\frac{\partial c}{\partial t}=-v_i\frac{\partial c}{\partial x}+D\frac{\partial^2c}{\partial x^2},
\end{equation}
where $v_i$ is determined by Eq.~(2), and  $D$ denotes the diffusion coefficient of the material. In the stationary state, the concentrations $C_{i-1}$ and $C_{i}$ at the inlet and outlet of the $i$th generation duct satisfy the equation:
\begin{equation}
C_{i-1}=\frac{A_i}{v_i}(1-e^{-v_ib_i/D})+C_{i}e^{-v_ib_i/D},
\end{equation}
where $A_i$ is the diffusion flux $A_i=v_ic-D\partial c/\partial x$ per unit area of a cross section of the duct. Because of the law of conservation of total diffusion flux at the connections of the parent and two daughter ducts, $\pi a_i^2A_i$ is equal to $2\pi a_{i+1}^2A_{i+1}$, that is, $A_i=1/(2r^2)^{i-1}A_1$. When the pressure difference $P=0$, the flow velocity $v_i=0$, and the material is transported only by diffusion. In this case, the concentration $C_i$ is explicitly expressed as  
\begin{equation}
C_i=C_{i-1}-\frac{b_i}{D(2r^2)^{i-1}}=C_{i-1}-\frac{bA_1}{D(2r)^{i-1}}=C_0-\frac{bA_1}{D}\frac{1-1/(2r)^{i}}{1-1/(2r)}.
\end{equation}
If $C_n$ is sufficiently small and $n$ is sufficiently large, $A_1$ is approximately expressed as $A_1=C_0D\{1-(1/2r)\}/b$. 
   
We assume that there are consumers of material just outside the deepest branch tips of generation $n$. In bronchial trees, the deepest branch tips correspond to the alveoli and the consumers correspond to the capillary vessels surrounding the alveoli. 
We further assume a reaction-diffusion equation in the consumers 
\begin{equation}
\frac{d C}{d t}=D^{\prime}(C_{n}-C)-\frac{kC^2}{1+KC^2},
\end{equation} 
where $C_{n}$ is the concentration of the outlet of the deepest $n$th duct, 
 and $D^{\prime}$ denotes the diffusivity between the deepest tips and consumers; (that is between the alveoli and the capillary vessels). In Eq.~(8), the Hill type reaction equation is assumed, taking the sigmoidal absorption kinetics of oxygen in capillary vessels into consideration. In the stationary state of Eq.~(8),  
\begin{equation}
D^{\prime}(C_{n}-C)=\frac{kC^2}{1+KC^2}=A_{n}=\frac{A_1}{(2r^2)^{n-1}}.
\end{equation}
It is noted that the reaction rate $kC^2/(1+KC^2)$ approaches a constant value $k/K$ for sufficiently large $C$ in the sigmoidal kinetics. 
The inlet concentration $C_0$ of the first duct is given  as the boundary condition. By solving the coupled equations Eqs.~(8) and (9), $A_1$ is determined, and the total reaction rate $S$ is defined as 
\begin{equation}
S=2^{n-1}\pi a_n^{2}A_n=\pi a^2A_1.
\end{equation}
Here, we propose a principle that the total reaction rate $S$ should be maximized in the optimal tree network. 

\begin{figure}[t]
\begin{center}
\includegraphics[height=4.cm]{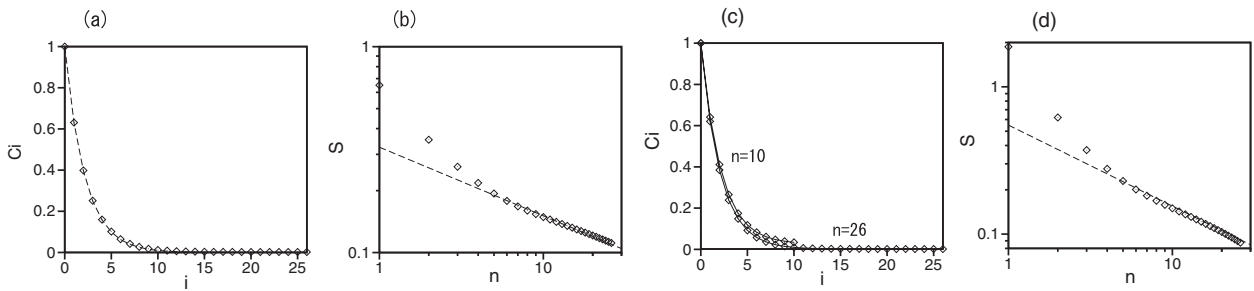}
\end{center}
\caption{Numerical results for  $k=100,K=1,D=D^{\prime}=0.1,\eta=1$ and $C_0=1$. (a) Profiles of $C_i$ for $n=26$ at $V_n=23,r=(1/2)^{1/3}$, and $P=0$. The dashed curve is calculated using Eq.~(7). (b) Relation of $n$ and $S$ for the parameter set of (a). The dashed line denotes $S\propto 1/n^{1/3}$.  
(c) Profiles of $C_i$ for $n=10$ and 26 at $V_n=23,r=0.81$, and $P=0.1$. (d) Relationship between $n$ and $S$ for the parameter set of (c). The dashed line denotes $S\propto 1/n^{0.55}$.  }
\label{f2}
\end{figure}
\begin{figure}[t]
\begin{center}
\includegraphics[height=4.cm]{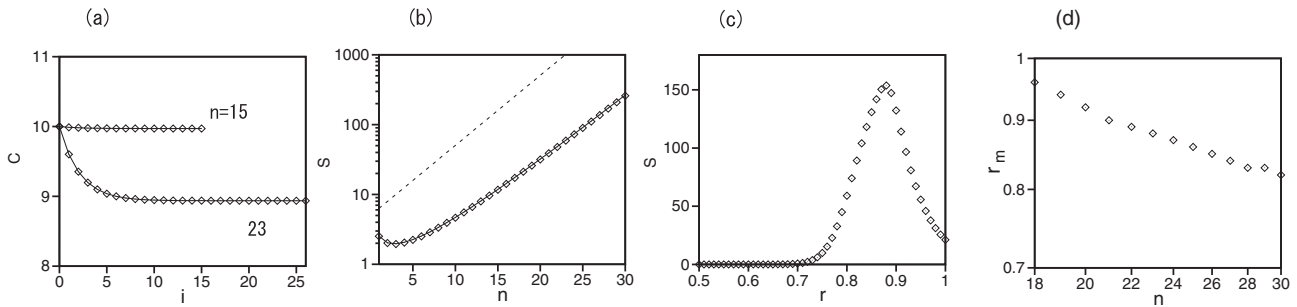}
\end{center}
\caption{Numerical results for $k=0.1,K=1,D=100,D^{\prime}=0.1,\eta=1,P=10$ and $C_0=10$. (a) Profiles of $C_i$ for $n=15$ and $26$ at $V_n=23$ and $r=0.8$. (b) Relationship between $n$ and $S$ for the same parameter set. The solid line denotes $S=\pi a^2(2r^2)^{n-1}(k/K)$ at $r=0.8$. The dashed line denotes $S\propto \exp\{(\log 2/3)n\}$. (c)  Relationship between $S$ and $r$ for $n=23$. (d) Ratio $r_m$ obtained by the maximization as a function of $n$.}
\label{f3}
\end{figure}

Although we performed numerical simulations for many parameter values, we show numerical results for some typical values. 
These parameter values correspond to transport-limited conditions, reaction-limited conditions of small and large Peclet numbers. The Peclet number is explained in a later paragraph. Since the aim of this paper is to study basic properties of our simple model, the parameter values do not always correspond to physiological conditions. The last conditions of large Peclet numbers correspond to physiological conditions of mammalian lungs qualitatively. 

Figure 2 shows numerical results for $k=0.1,K=1,D=100,D^{\prime}=0.1,\eta=1,P=10$ and $C_0=10$. Figure 2(a) is a profile of $C_i$ (rhombi) for $n=26$ at $V_n=23,r=(1/2)^{1/3},$ and $P=0$. In this case, the pressure difference is zero and the flow velocity is zero. The concentration $C_i$ decreases with $i$ monotonically. The dashed line is calculated using Eq.~(7), where $A_1$ is approximated by $A_1=C_0D\{1-(1/2r)\}/b$. Good agreement is observed between the theory and the numerical results. The total reaction rate $S$ is expressed as $S=\pi a^2 A_1\sim \pi a^2/b C_0\{1-(1/2r)\}$ for sufficiently large $n$.
At $r=(1/2)^{1/3}$, $a\propto 1/n^{1/3}$ and $b\propto 1/n^{1/3}$. Therefore, $S\propto 1/n^{1/3}$. 
Figure 2(b) shows the relationship between $n$ and $S$ obtained from the numerical simulation of the same parameter set. As predicted by the theory, $S\propto 1/n^{1/3}$ is satisfied for large $n$. In the present case of a diffusion-limited reaction, the total reaction rate decreases monotonically with $n$. That is, the complicately branched network is not efficient for the reactions at the deepest branch tips, because the transport of material  by diffusion limits the total reaction rate. Next, we investigated the effect of the flow induced by a  pressure difference.  Figure 2(c) shows the profiles of $C_i$ for $n=10$ and 26 at $V_n=23,r=0.81$, and $P=0.1$. A small pressure difference $P=0.1$ is added and the ratio $r$ is changed slightly. The concentration $C_i$ decreases rapidly with $i$ toward zero at $n=26$. Figure 2(d) shows the relationship between $n$ and $S$ for the same parameter set. The dashed line $S\propto 1/n^{0.55}$ is obtained by the curve-fitting for large $n$. The total reaction rate $S$ decreases with $n$ more rapidly than the case of Fig.~2(b). The advection term is added but the transport of material is still a limiting factor for the reaction. These numerical results suggest that the complicatedly branched network with a large $n$ is not useful in the principle of maximization of the total reaction rate $S$ in a system under transport-limited conditions. 

Figure 3 shows numerical results for $k=0.1,K=1,D=100,D^{\prime}=0.1,\eta=1,P=10$ and $C_0=10$. Figure 3(a) shows the profiles of $C_i$ for $n=15$ and $26$ at $V_n=23$, and $r=0.8$. The concentration is high even for large $i$. This is because the transport is sufficiently fast owing to the large values of parameters $D$, $P$, and $C_0$, and the reaction is slow  owing to the small value of parameter $k$.  The reaction rate $A_n$ outside the deepest branch tips is expected to saturate at large values of concentration.    Figure 3(b) shows the relationship between $n$ and $S$ for the same parameter set on a semi-logarithmic plot. The total reaction rate $S$ increases exponentially with $n$ for $n>5$. The reaction outside the deepest branch tips is the rate-limiting process at this parameter set. In this case, the total reaction rate increases with the number of total reaction sites, which increases exponentially with the branching number $n$. If $KC^2$ is sufficiently larger than 1, $A_n=k/K$ is a good approximation owing to the saturation effect of the reaction as is noted below Eq.~(9). Therefore, $S$ is evaluated as $S=2^{n-1}\pi a_n^2 A_n=\pi a^2(2r^2)^{n-1}(k/K).$ 
The solid line in Fig.~3(b) shows the relation $S=\pi a^2(2r^2)^{n-1}(k/K)$ for $r=0.8$. It should be noted here, that $a$ is a function of $n$ owing to the condition that $V_n=$const. Very good agreement between numerical results denoted by rhombi and the theoretical prediction denoted by solid line is seen in Fig.~3(b). The relationship between $S$ and $n$ approaches $S\propto 2^{(n-1)/3}=\exp\{(\log 2/3)n\}$ for sufficiently large $n$.  Figure 3(c) shows the relationship between $S$ and $r$ for $n=23$  and  $V_n=23$. The total reaction rate $S$ reaches a maximum at $r=0.88$ for the generation number $n=23$.  This is definitely larger than Murray's ratio $r=(1/2)^{1/3}$.  We have calculated the ratio $r_m$ by maximizing $S$ for various values of $n$ for the same parameter set. Figure 3(d) shows the ratio $r_m$ obtained by the maximization of $S$ as a function of $n$. These results suggest that Murray's ratio is not always obtained, although the ratio $r_m$ decreases with $n$ and tends to approach  Murray's ratio for large $n$.    

\begin{figure}[t]
\begin{center}
\includegraphics[height=4.cm]{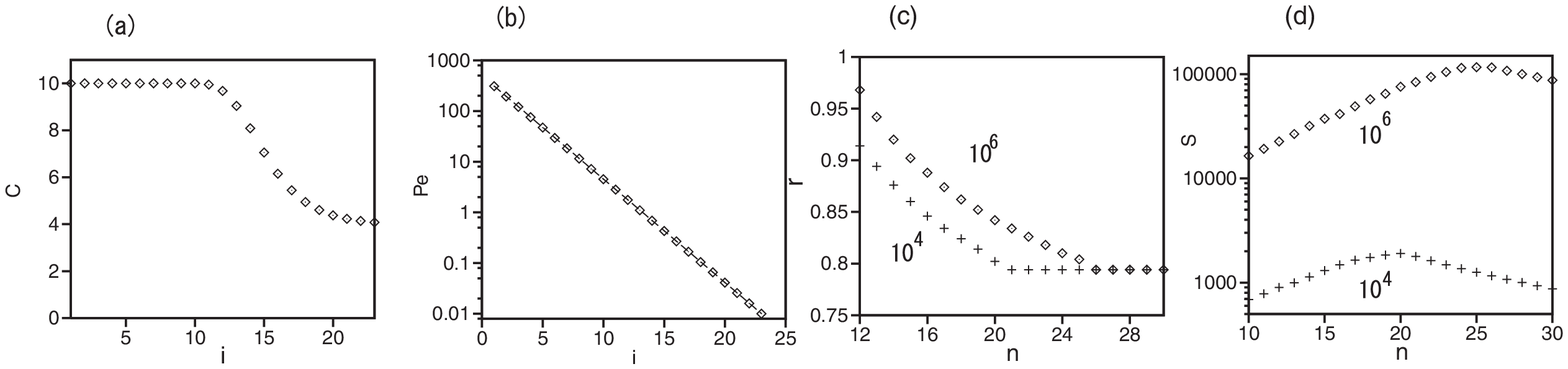}
\end{center}
\caption{Numerical results for $k=0.1,K=1,D=D^{\prime}=0.1,\eta=1,P=10$ and $C_0=10$. (a) Profiles of $C_i$ for $n=23$ at $V_n=10^6$ and $r=0.818$. (b) Relationship between $i$ and $Pe$ for the same parameter set. The dashed line is $Pe=500e^{-0.47i}$. (c) Relationship between the optimal size ratio $r$ and $n$ at $V_n=10^6$ (rhombi) and $V_n=10^4$ (plus). (d) Relationship between the maximum reaction rate $S_m$ and $n$ at $V_n=10^6$ (rhombi) and $V_n=10^4$ (plus).}
\label{f4}
\end{figure}
In the numerical simulations shown in Figs.~2 and 3, the convection term was not  important. The effect of convection can be evaluated by the Peclet number, which is defined as $Pe=v_ib_i/D$. It is an important non-dimensional parameter. The advection term in Eq.~(5) is dominant when $Pe>1$, and the diffusion term is dominant when $Pe<1$. The Peclet number decreases exponentially with $i$ as $Pe\propto  1/(2r)^{i-1}$ when $r>1/2$. In the deeper generations of ducts, the diffusion term becomes dominant, because the Peclet number decreases exponentially. In human lungs, the Peclet number is $O(10^3)$ at $i=1$. In the  numerical simulations shown in Figs.~2 and 3, the Peclet number is smaller than 1 even at $i=1$. 

We investigate systems of larger Peclet number, keeping human lungs in mind. We show some numerical results at larger values of $V_n$,  because $v_i$ and $b_i$ in the Peclet number $v_ib_i/D$ increase with $V_n$ owing to Eqs.~(1) and (2).   Figures 4(a) and (b) are numerical results for $n=23$ at $V_n=10^6,k=0.1,K=1,D=D^{\prime}=0.1,\eta=1,P=10,C_0=10$ and $r=0.818$. 
Figure 4(a) is a profile of the concentration $C_i$. The concentration $C_i$ remains constant at $C_0=10$ for $i\le 12$. This flat profile of $C_i=C_0$ for $i\le 12$ is due to the large Peclet number. $C_i$ decreases monotonically for $i>12$, however, the concentration $C_n$ at the deepest branch tips is still around 4, and the saturation condition $KC^2>>1$ of the reaction is satisfied. 
That is, the reaction-limited condition is satisfied at this parameter set. Figure 4(b) shows a semi-logarithmic plot of the Peclet number $Pe$ as a function of $i$. The Peclet number decreases exponentially with $i$ as $Pe=500e^{-0.47 i}$. We have calculated the optimal size ratio $r$ by maximizing the total reaction rate $S$ for various fixed values of $n$. 
Figure 4(c) shows the relationship between the optimal size ratio $r$ and $n$ at $V_n=10^6$ and $10^4$ respectively. 
The optimal size ratio decreases with $n$, and approaches the value $(1/2)^{1/3}\sim 0.794$. Figure 4(d) shows the relationship between the maximum reaction rate $S_m$ and $n$. $S_m$ reaches a maximum value at $n_c=20$ for $V_n=10^4$ and $n_c=25$ for $V_n=10^6$. 
The optimal values $n_c$ of the branching number $n$ correspond to the critical points just before the optimal size ratio $r$ changes to the constant value 0.794 as shown in Fig.~4(c).  For $n>n_c$, the concentration $C_n$ at the deepest branch tips becomes sufficiently small. The system changes from the reaction-limited conditions to the transport-limited conditions near $n=n_c$. As a result, the total reaction rate $S_m$ decreases with $n$ for $n>n_c$ as shown in Fig.~4(d).  The maximization of the total reaction rate $S$ is satisfied at the maximization of the flow rate $Q$ under this convective-transport limited condition in the case of large Peclet numbers. This is the reason why the optimal size ratio takes a constant value $r=(1/2)^{1/3}$ for $n>n_c$ in Fig.~4(c).  In any case, the size ratio $r=(1/2)^{1/3}$ is obtained as a result of the doubly optimal conditions with respect to the size ratio $r$ and the branching number $n$.    

In summary, we have proposed a simple reaction-diffusion-advection equation in a dichotomous tree network.  An optimal size ratio $r$ is evaluated by the principle of  maximization of  total reaction rate. In case of reaction-limited conditions, the dichotomous tree network is efficient, and the optimal ratio can be larger than $(1/2)^{1/3}$ under the condition of fixed branching $n$, which is consistent with observations in mammalian lungs. Furthermore, we have found that there is an optimal generation number $n_c$ when the Peclet number is large.  For the optimal branching number $n_c$, the optimal size ratio is close to $r=(1/2)^{1/3}$. Our simple model might be applied as rough approximation for various network systems. More realistic models of human lungs will be topics of future research studies.

\end{document}